\documentclass[a4paper,11pt]{article}
\usepackage{pos}
\usepackage{physics}

\title{Universal scaling and the asymptotic
behaviour of Fourier coefficients
of the baryon-number density in QCD}
\ShortTitle{Universal scaling and Fourier coefficients}

\author*[a]{C. Schmidt}
\author[b]{D. A. Clarke}
\author[c]{P. Dimopoulos}
\author[c]{F. Di Renzo}
\author[d]{J. Goswami}
\author[a]{S. Singh}
\author[e]{V. V. Skokov}
\author[f]{K. Zambello}

\affiliation[a]{Universität Bielefeld, Fakultät für Physik, Universitätsstrasse 25, 33615 Bielefeld, Germany}

\affiliation[b]{Department of Physics and Astronomy, University of Utah,
Salt Lake City, Utah  84112, United States}

\affiliation[c]{Dipartimento di Scienze Matematiche, Fisiche e Informatiche, 
Università di Parma and INFN, Gruppo Collegato di Parma 
I-43100 Parma, Italy}

\affiliation[d]{RIKEN Center for Computational Science,
Kobe 650-0047, Japan}

\affiliation[e]{Department of Physics, North Carolina State University,
Raleigh, NC 27695, United States}

\affiliation[f]{Dipartimento di Fisica dell’Università di Pisa and INFN--Sezione di Pisa,
Largo Pontecorvo 3, I-56127 Pisa, Italy.}

\newcommand{\TRW}{T_{\rm RW}}
\newcommand{\R}{\mathbb{R}}
\newcommand{\simulat}{\texttt{SIMULATeQCD}}

\emailAdd{schmidt@physik.uni-bielefeld.de}

\abstract{We discuss the scaling of the Yang-Lee singularity (YLs) and show how the universal scaling can be used to locate phase transitions in QCD. We describe two complementary methods to extract the location of the Yang-Lee singularity from lattice QCD data of the baryon-number density and higher order cumulants of the baryon number, obtained at imaginary chemical potential. The first method (multi-point Pad\'e resummation) is used to determine the Roberge-Weiss phase transition temperature. Our continuum extrapolated result is $\TRW=211.1\pm3.1$ MeV. The second method is based on the asymptotic behaviour of the Fourier coefficients of the baryon-number density. We discuss the derivation of a fitting function and demonstrate that the procedure can successfully locate the YLs in the Quark Meson model.}

\FullConference{The 40th International Symposium on Lattice Field Theory (Lattice 2023)\\
July 31st - August 4th, 2023\\
Fermi National Accelerator Laboratory\\}

\begin{document}
\maketitle

\section{Introduction}
The QCD phase diagram is subject to active  research since the introduction of the quark model. It is of utmost importance for cosmology, astrophysics and heavy ion phenomenology. Establishing the existence and location of the QCD critical point at nonzero temperature $T$ and baryon chemical potential $\mu_B$ is considered a grand challenge in lattice QCD. Unfortunately, lattice QCD is hampered by the infamous sign problem such that direct lattice QCD calculations at $\mu_B>0$ are impractical. We thus have to rely on extrapolation methods based on Taylor expansion \cite{Allton:2002zi} or analytic continuations from calculations at imaginary chemical potentials $\mu_B=i\theta$, $\theta\in\R$ \cite{DElia:2002tig,deForcrand:2002hgr}. These methods are in principle sensitive to singularities in the complex chemical potential plane \cite{Bollweg:2022rps}. In particular, if the information from lattice QCD data is converted into a rational function form, it is straightforward to determine the roots of the denominator and thereby extract the positions of singularities.

In lattice QCD calculations at $T>0$, and statistical physics in general, derivatives of the logarithm of the partition function $Z$ are studied in order to investigate phase transitions. The scaling hypothesis dictates that these observables change with the reduced temperature ($t$) or symmetry breaking field parameter ($h$) as a power law, exposing the universal critical exponents. Thus sufficiently high derivatives diverge at $(t,h)=(0,0)$. In a finite volume, $Z$ can develop a countable number of zeros, or equivalently singularities of $\ln Z$. It is well known that for $t>0$ the zeros move into the complex $h$-plane in a particular manner \cite{Yang:1952be}. The Yang-Lee singularity (YLs), which is approximated by the first zero (up to finite size effects),  moves along the imaginary $h$-axis and scales with temperature such that it takes a universal (constant) position in the scaling variable $z=t/h^{1/\beta\delta}\equiv z_c$. Here $\beta,\delta$ are critical exponents of the underlying universality class. The universal position of the YLs has been determined only very recently \cite{Connelly:2020gwa}.  

Here we use derivatives of $\ln Z$ w.r.t.~$\mu_B$ as our input data. These observables characterize fluctuations of the conserved (net) baryon number, i.e. higher derivatives are the cumulants of the baryon-number distribution. It has been realized that tracking the YLs in the complex $\mu_B$-plane, which have been determined from the baryon-number density, is a novel and possibly very robust method for the determination of phase transition in QCD \cite{Dimopoulos:2021vrk,Singh:2023bog,FDR,CEP,Goswami:2024jlc,Basar:2023nkp}. We will discuss two methods that are used to extract the YLs from the lattice QCD data. The first method is the multi-point Pad\'e method, which has been introduced in \cite{Dimopoulos:2021vrk}. Using this method we will report on the status of the determination of the Roberge-Weiss transition temperature $\TRW$ in QCD, extracted through the universal scaling of the YLs. Applying this method also in the vicinity of the QCD critical end point is very tempting and will be discussed elsewhere \cite{CEP,Goswami:2024jlc}. For a discussion of finite size scaling using the multi-point Pad\'e approach, see \cite{Singh:2023bog,FDR}. As a second procedure to extract YLs from the lattice data, we will introduce and discuss the asymptotic behavior of the Fourier coefficients of the baryon-number density. We show that the first $\order{20}$ Fourier coefficients might be sufficient to extract the position of the YLs, given the data is sufficiently precise. We demonstrate the efficiency of the method in the Quark-Meson Model \cite{Fourier}. 

\section{The multi-point Pad\'e method}
\begin{figure}
    \centering
    \includegraphics[width=0.49\textwidth]{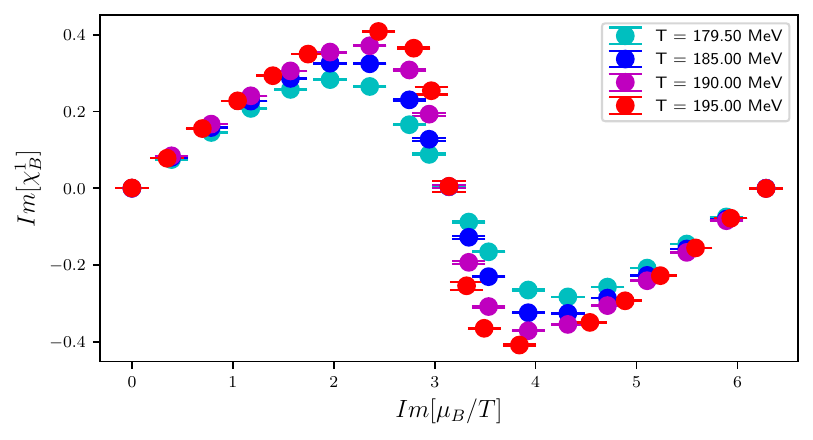}
    \includegraphics[width=0.49\textwidth]{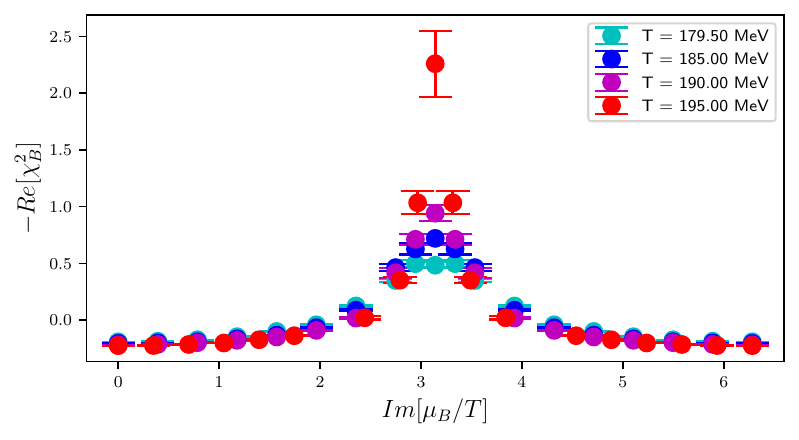}
    \caption{First- and second-order cumulants of the baryon number as a function of the imaginary chemical potential, calculated on $36^3\times 6$ lattices. }
    \label{fig:data}
\end{figure}

\begin{figure}
    \centering
    \includegraphics[width=0.49\textwidth]{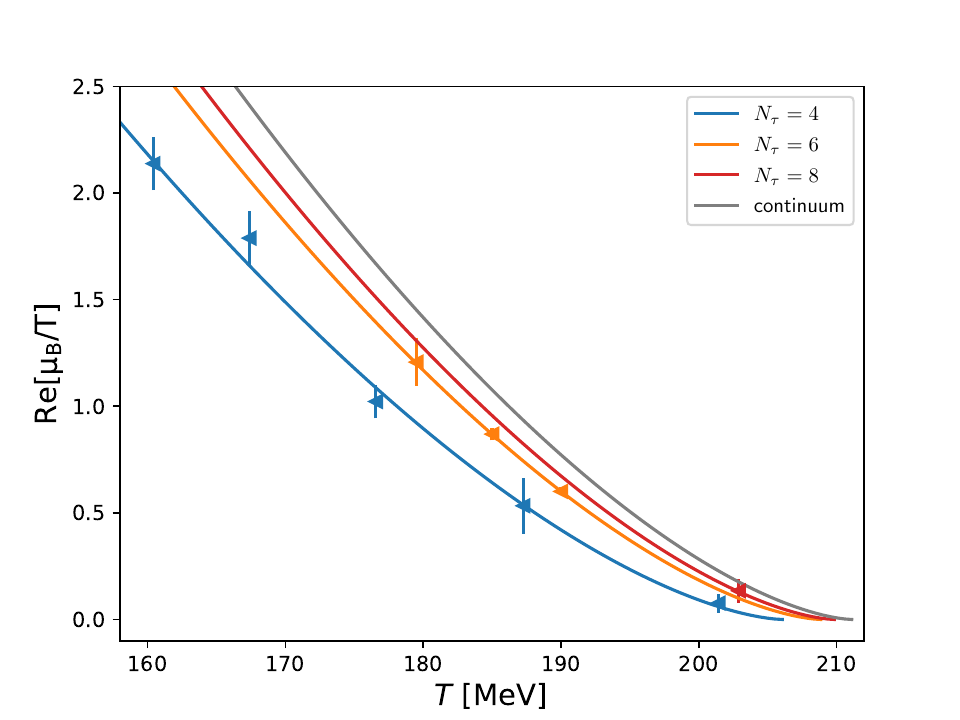}
    \includegraphics[width=0.49\textwidth]{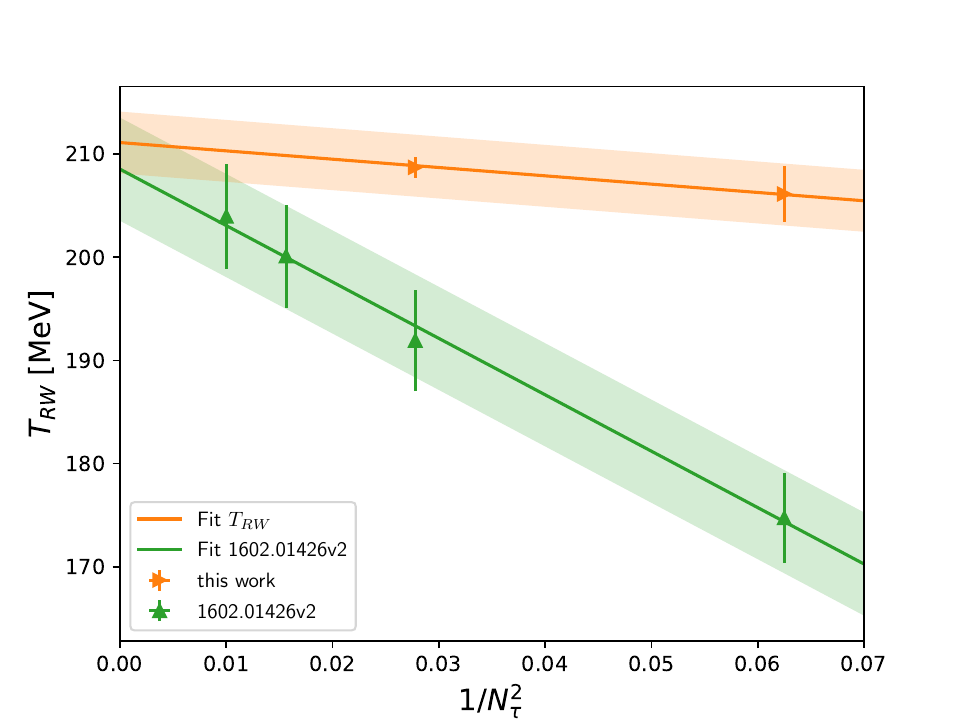}
    \caption{Scaling of the Yang-Lee singularity with temperature for three different lattice spacings ($N_\tau$) and continuum extrapolation (left). The $N_\tau$ dependence of the Roberge-Weiss transition temperature $\TRW(N_\tau)$ from this work and from \cite{Bonati:2016pwz}. }
    \label{fig:TRW}
\end{figure}

We perform lattice QCD calculations with the highly improved staggered quark (HISQ) action using the $\simulat$ software \cite{Altenkort:2021cvg,HotQCD:2023ghu}. We are using (2+1) flavors with physical masses, using the scale setting procedure and line of constant physics as described by the HotQCD collaboration \cite{Bollweg:2021vqf}. The observables we determine are the cumulants of the (net) baryon number, given as 
\begin{equation}\begin{aligned}
 \chi_n^B(T,V,\mu_B) &= \frac{1}{VT^3} \left( \frac{\partial}{\partial \hat{\mu}_B} \right)^n ~ \ln ~ Z(T, V, \mu_l, \mu_s) \\
& =  \frac{1}{VT^3} \left( \frac{1}{3} \frac{\partial}{\partial \hat{\mu}_l} + \frac{1}{3} \frac{\partial}{\partial \hat{\mu}_s} \right)^n \ln ~ Z(T, V, \mu_l, \mu_s) \mbox{ . }
\end{aligned}\end{equation}
Here $\mu_l$ and $\mu_s$ are the chemical potentials of the two light flavors and of the strange quark, respectively. We also have introduced the dimensionless notation $\hat\mu=\mu/T$. For simplicity we use $\mu_l=\mu_s$, which corresponds to $\mu_B=3\mu_l$ and zero strangeness chemical potential $\mu_S=0$\footnote{For the conversion of the quark chemical potentials $\mu_u, \mu_d, \mu_s$ to the hadronic chemical potential $\mu_B, \mu_Q,\mu_S$ see e.g. Ref.~\cite{HotQCD:2012fhj}. }. The derivatives are expressed in space-time traces over operators, which are appropriate combinations of the inverse and the derivatives of the fermion matrix. We calculate the traces by the random noise method, using normally  distributed volume sources. Moments of the traces are calculated in an unbiased manner \cite{Mitra:2022vtf}.
For temperatures $T\lesssim 200$ MeV, we calculate on lattices of size $N_\sigma^3\times N_\tau=24^3\times 4$, $36^3\times 6$ and $48^3\times 8$. The results for the first two cumulants from the $N_\tau=6$ lattices are shown in Fig.~\ref{fig:data}. The statistics varies between 2000 and 6000 configurations per simulation point $(\mu_B,T)$.
From the data we determine the multi-point Pad\'e of $\chi_1^B$ in $\theta=\text{Im}\hat\mu_B$ for each temperature $T$ by the ansatz 
\begin{equation}
    \chi_1^B=R^m_n(\theta)=\frac{\sum_{i=0}^ma_i\theta^i}{1+\sum_{j=1}^n b_j\theta^j}\;. 
\end{equation}
The coefficients $a_i,b_i$ are determined by solving a set of linear equations, making use of $\chi_1^B$ and $\chi_2^B$ data \cite{Dimopoulos:2021vrk}.
Finally, we calculate the roots of the denominator  to obtain the closest (uncanceled) singularity $\hat\mu_{YL}$, which approximates the position of the YLs in the infinite volume limit. 
To apply Yang-Lee scaling, we have to express the universal scaling fields $t,h$ in terms of the QCD parameters $T,\mu_B$. In the vicinity of the Roberge-Weiss transition \cite{Roberge:1986mm}, located at $\hat\mu_B=i\pi$, we have the following relations 
\begin{equation}
    t=t_0^{-1}\frac{\TRW-T}{\TRW}, \qquad h=h_0^{-1}\frac{\hat\mu_B-i\pi}{i\pi},
\end{equation}
where $t_0, h_0$ are normalization constants and $\TRW$ is the Roberge-Weiss phase transition temperature. Note that the imaginary direction of the symmetry breaking field is mapped onto the real direction in the chemical potential $\hat\mu_B$. We can now solve $t/h^{1/\beta\delta}=z_c$ to obtain $\hat\mu_{YL}(T)$. Here we use the ansatz
\begin{equation}
    \hat\mu_{YL}(T)=a(N_\tau)\left(\frac{\TRW(N_\tau)-T}{\TRW(N_\tau)}\right)^{\beta\delta},
\end{equation}
where we assume the following cut-off dependence of the Roberge-Weiss transition and the amplitude:  $\TRW(N_\tau)=\TRW^{(0)}+\TRW^{(2)}/N_\tau^2$ and $a(N_\tau)=a^{(0)}+a^{(2)}/N_\tau^2$. We thus have four fit parameters in total. In Fig.~\ref{fig:TRW} (left) we show that the ansatz works well for the available data ($\chi^2\approx 0.4$).
We obtain the continuum result $\TRW^{(0)}=211.1\pm 3.1$ MeV. A comparison with the result from \cite{Bonati:2016pwz} is shown in Fig.~\ref{fig:TRW} (right). Note that the two continuum extrapolations are in good agreement. 

\section{The Fourier coefficients}
\begin{figure}
    \centering
    \includegraphics[width=\textwidth]{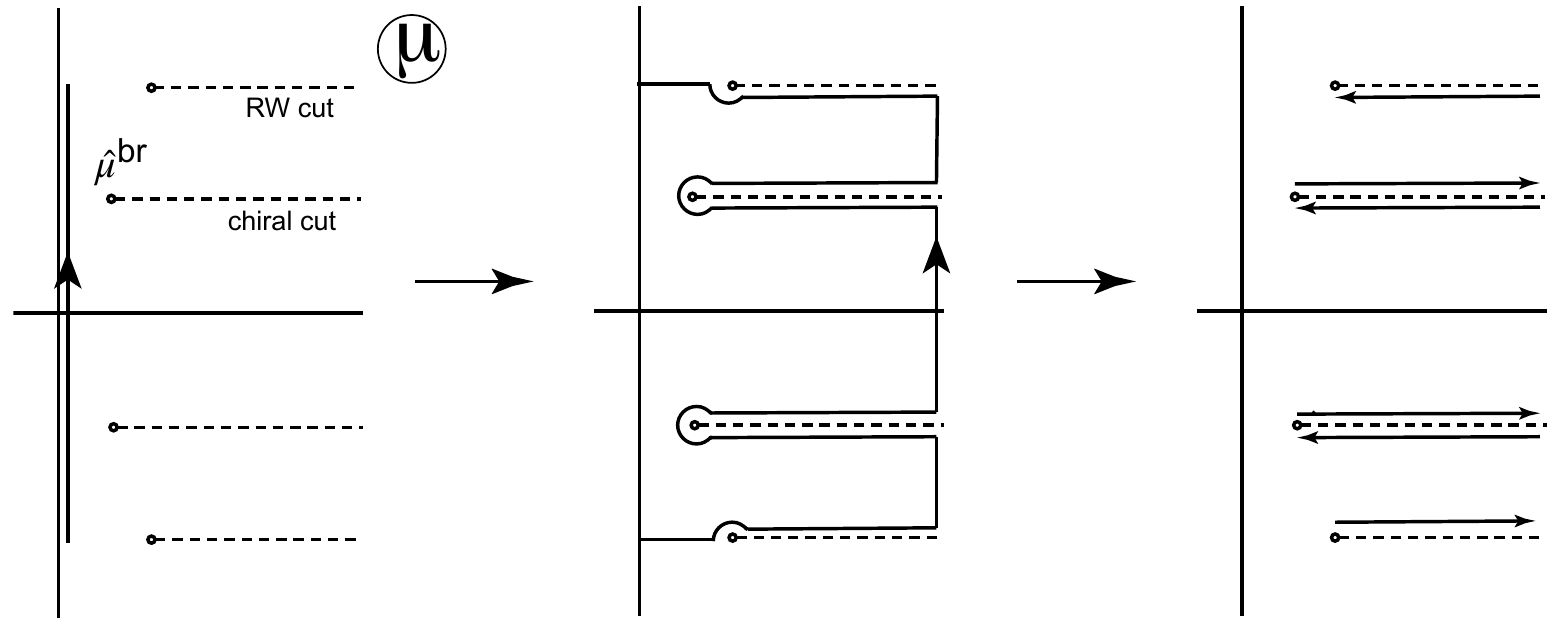}
    \caption{Deformation of the integration contour into the complex $\hat\mu_B$ plane along the Roberge-Weiss (thermal) cuts as well as the Yang-Lee cut of the chiral transition. }
    \label{fig:contour}
\end{figure}

A complementary approach to the YLs is related to the Fourier coefficients $b_k$ of the baryon-number density $\hat n_B=n_B/T^3=\chi_1^B$, defined as 
\begin{align}
    \label{Eq:bk}
    b_k = \frac{1}{i \pi} \int_{-\pi}^{\pi} d\theta  \,  \hat n_B (\hat \mu = i \theta ) 
    \sin \left(k \theta \right).
\end{align}
In order to evaluate the integral in Eq.~(\ref{Eq:bk}) we deform the integration contour into the complex plane. This is valid as long as we do not cross any non-analyticities. It is important to notice that the YLs condense in the continuum limit to form branch-cuts in $\hat n_B$. We thus deform the integration contour to follow those branch cuts as shown in Fig.~\ref{fig:contour}. We indicated here the branch cuts of the Roberge-Weiss (thermal) transition as well as the ones of the chiral phase transition. Note that they come in complex conjugate pairs. After the deformation, the first and the last segment cancel each other due to the periodicity in $\hat\mu_B$. The segments at infinity vanish due to the exponential decay of the integrand. 
Finally only the segments along the cuts remain. 

As the next important step we express the baryon-number density by its universal behaviour in the vicinity of the branch point $\hat\mu\to\hat\mu^{\rm br}$,
\begin{equation}
    \hat n_B(\hat \mu)  =  A (\hat \mu - \hat \mu^{\rm br})^{\sigma} 
(1 + B(\hat \mu - \hat \mu^{\rm br})^{\theta_c} + \ldots )  + 
\sum\limits_{n=0}^{\infty} a_n (\hat \mu - \hat \mu^{\rm br})^n,
\end{equation}
where $\sigma$ denotes the leading order Yang-Lee edge critical exponent and $\theta_c$ the confluent critical exponent. The last term represents the regular contribution. It is interesting to note that the regular contribution exactly vanishes on the deformed integration contour. After injecting this expression for $\hat n_B$ into Eq.~(\ref{Eq:bk}) and performing the integrals, we obtain \cite{Fourier} 
\begin{align}
    b_k = 
      \tilde A_{\rm YLE} \frac{e^{-\hat \mu^{\rm YLE} k }}{k^{1+\sigma}} \left(
    1  +   \frac{\tilde B_{\rm YLE}}{k^{\theta_c}}  
    + \ldots \right)
    + \tilde A_{\rm RW} \frac{e^{-\hat \mu^{\rm RW} k }}{k^{1+\sigma}} \left(
    1  +   \frac{\tilde B_{\rm RW}}{k^{\theta_c}}  
   + \ldots \right) + {\rm c.c.}\,.
\end{align}
Dropping the sub-leading contribution and noting that $\rm Im\mu^{RW}=\pi$, results in
\begin{align}
\label{Eq:fit}
    b_k &= 
      |\tilde A_{\rm YLE}| \frac{e^{-\hat \mu_r^{\rm YLE} k }}{k^{1+\sigma}}
    \cos(\hat \mu_i^{\rm YLE} k + \phi^{\rm YLE}_a) 
    + |\hat A_{\rm RW}|  (-1)^k \frac{e^{-\hat \mu_r^{\rm RW} k }}{k^{1+\sigma}},
\end{align}
which can be used as a fitting function to extract the position of the branch-cut singularity. Some remarks are in order: The exponential decay of the coefficients determines the real part of the branch-cut singularity $\mu_r$, while the imaginary part $\mu_i$ is indicated by the oscillations. We usually assume that only one branch-cut singularity dominates, the one which has the smaller real part $\mu_r$. The Roberge-Weiss branch cut is located at $\hat\mu_i^{RW}=\pi$ such that the oscillations are maximally $(-1)^k$, as long as the Roberge-Weiss branch cut is dominating. 

\begin{figure}
    \centering
        \includegraphics[width=0.49\linewidth]{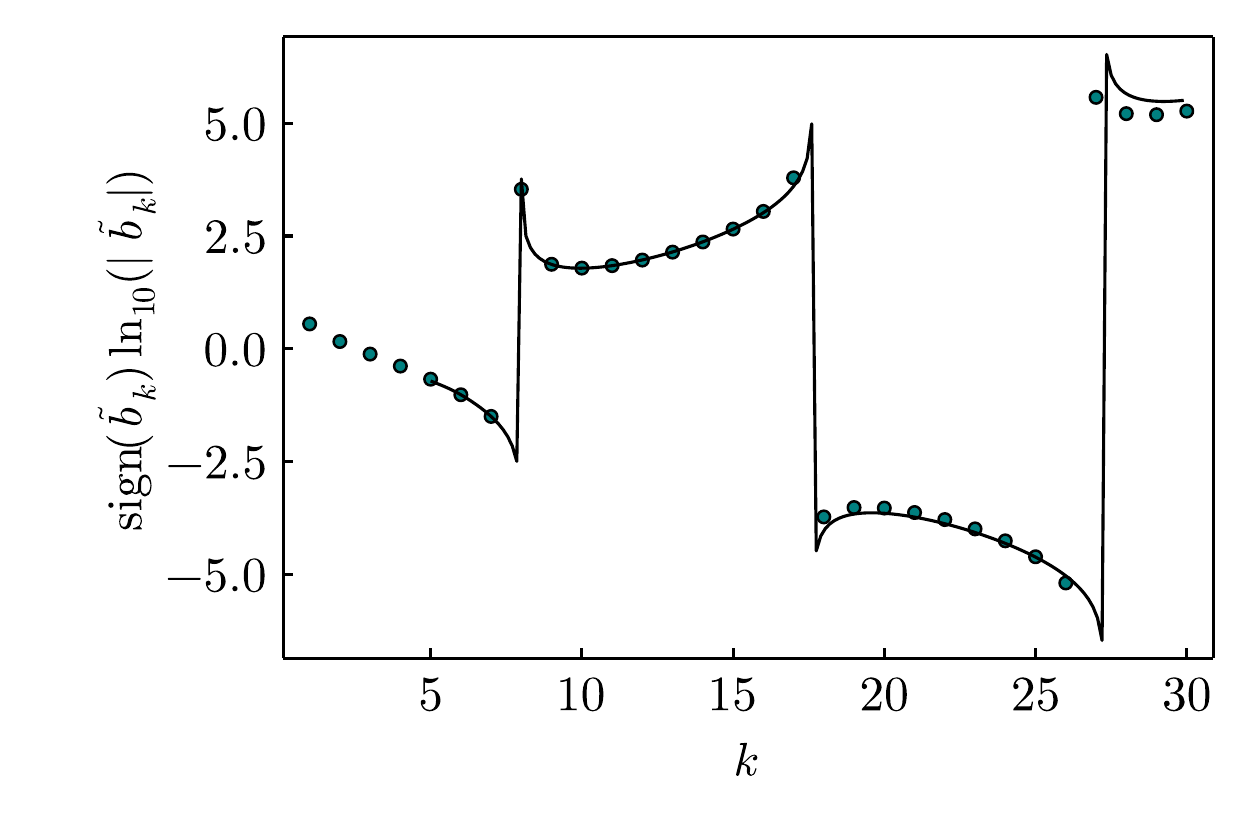}
    \includegraphics[width=0.49\linewidth]{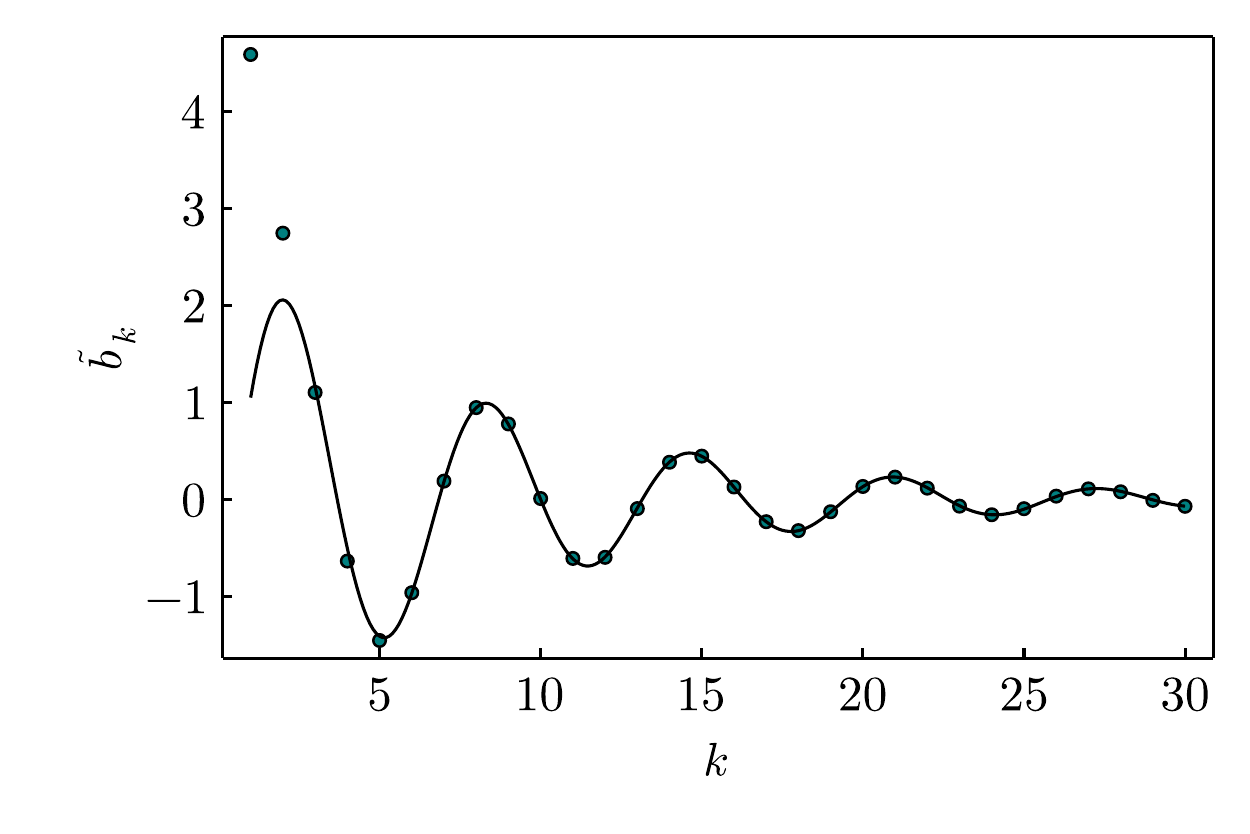}
    \caption{Mean-field Fourier coefficients $\tilde b_k = k^{1+\sigma} b_k$ ($\sigma_{\rm MF} = 1/2$) for $T = 150$ MeV (left) and $T = 180$ MeV (right)   and the corresponding fits.}
    \label{fig:MF}
\end{figure}

In Fig.~\ref{fig:MF} we apply this fitting function to Fourier coefficients obtained in the mean-field approximation of the Quark Meson model at $T=150$ and $T=180$ MeV. Details of the model are given in Ref.~\cite{Skokov:2010sf}.
We see that the ansatz works very well and that the asymptotic behaviour already sets in at $k\gtrsim5$. The fit reproduces the exact position of the branch-cut singularity within 7\% precision. 

We will apply this analysis to lattice QCD data in the  future. The challenge is the precise calculation of the Fourier coefficients. To tackle this challenge we are currently exploring asymptotically correct quadrature rules designed for highly oscillatory integrals. 

\section*{Acknowledgement}
This work is supported by the
Deutsche Forschungsgemeinschaft (DFG, German Research Foundation) - Project number 315477589-TRR
211, the PUNCH4NFDI consortium supported by
the Deutsche Forschungsgemeinschaft (DFG, German
Research Foundation) with project number 460248186
(PUNCH4NFDI) and by the European Union under grant agreement No. H2020-MSCAITN-2018-813942 (EuroPLEx). 
DAC is supported by the National Science Foundation under Grants PHY20-13064 and PHY23-10571.
FDR acknowledges support by INFN under the research program \textit{i.s.} QCDLAT.
VVS is supported by the U.S. Department of Energy, Office of Nuclear Physics through contract DE-SC0020081.

\end{document}